\documentclass[twocolumn,pra,showpacs]{revtex4-1}
\usepackage{graphicx,amsmath,amssymb,xspace}
\usepackage{color}

\begin{document}

\newcommand{\ketbra}[2]{| #1\rangle \langle #2|}
\newcommand{\ket}[1]{| #1 \rangle}
\newcommand{\bra}[1]{\langle #1 |}
\newcommand{\Tr}{\mathrm{Tr}}
\newcommand\F{\mbox{\bf F}}
\newcommand{\h}{\mathcal{H}}

\newcommand{\PSD}{\textup{PSD}}

\newcommand{\C}{\mathbb{C}}
\newcommand{\X}{\mathcal{X}}
\newcommand{\Y}{\mathcal{Y}}
\newcommand{\Z}{\mathcal{Z}}
\newcommand{\sspan}{\mathrm{span}}
\newcommand{\kb}[1]{\ket{#1} \bra{#1}}
\newcommand{\pos}{D}

\newcommand{\thmref}[1]{\hyperref[#1]{{Theorem~\ref*{#1}}}}
\newcommand{\lemref}[1]{\hyperref[#1]{{Lemma~\ref*{#1}}}}
\newcommand{\corref}[1]{\hyperref[#1]{{Corollary~\ref*{#1}}}}
\newcommand{\eqnref}[1]{\hyperref[#1]{{Equation~(\ref*{#1})}}}
\newcommand{\claimref}[1]{\hyperref[#1]{{Claim~\ref*{#1}}}}
\newcommand{\remarkref}[1]{\hyperref[#1]{{Remark~\ref*{#1}}}}
\newcommand{\propref}[1]{\hyperref[#1]{{Proposition~\ref*{#1}}}}
\newcommand{\factref}[1]{\hyperref[#1]{{Fact~\ref*{#1}}}}
\newcommand{\defref}[1]{\hyperref[#1]{{Definition~\ref*{#1}}}}
\newcommand{\exampleref}[1]{\hyperref[#1]{{Example~\ref*{#1}}}}
\newcommand{\hypref}[1]{\hyperref[#1]{{Hypothesis~\ref*{#1}}}}
\newcommand{\secref}[1]{\hyperref[#1]{{Section~\ref*{#1}}}}
\newcommand{\chapref}[1]{\hyperref[#1]{{Chapter~\ref*{#1}}}}
\newcommand{\apref}[1]{\hyperref[#1]{{Appendix~\ref*{#1}}}}
\newcommand\rank{\mbox{\tt {rank}}\xspace}
\newcommand\prank{\mbox{\tt {rank}$_{\tt psd}$}\xspace}
\newcommand\alice{\mbox{\sf Alice}\xspace}
\newcommand\bob{\mbox{\sf Bob}\xspace}
\newcommand\pr{\mbox{\bf Pr}}
\newcommand\av{\mbox{\bf{\bf E}}}
\newcommand{\pabxy}{(p(ab|xy))}
\newcommand{\calQ}{\mathcal{Q}}
\def\be{\begin{equation}}
\def\ee{\end{equation}}

\newcommand{\snote}[1]{\textcolor{blue}{\textbf{(Jamie: #1)}}}
\newcommand{\wnote}[1]{\textbf{[***#1***]}}
\newcommand{\comment}[1]{{}}
\newcommand{\NEW}[1]{\textcolor{blue}{#1}}
\newcommand{\red}[1]{\textcolor{red}{#1}}

\title{\vspace{-1cm} Device-independent characterizations of a shared quantum state independent of any Bell inequalities}

\author{Zhaohui Wei$^{1,2,3,}$}\email{Email: weizhaohui@gmail.com}
\author{Jamie Sikora$^{1,3}$}
\affiliation{$^{1}$Centre for Quantum Technologies, National
University of Singapore, Singapore\\$^{2}$School of Physical and
Mathematical Sciences, Nanyang Technological University,
Singapore\\$^{3}$MajuLab, CNRS-UNS-NUS-NTU International Joint
Research Unit, UMI 3654, Singapore}

%\date{August 31, 2016}

\begin{abstract}
In a Bell experiment two parties share a quantum state and perform
local measurements on their subsystems separately, and the
statistics of the measurement outcomes are recorded as a Bell
correlation. For any Bell correlation, it turns out {that} a quantum
state with minimal size that is able to produce this correlation can
always be pure. In this work, we first exhibit two
device-independent characterizations for the pure state that Alice
and Bob share using only the correlation data. Specifically, we give
two conditions that the Schmidt coefficients must satisfy, which can
be tight, and have various applications in quantum tasks. First, one
of the characterizations allows us to bound the entanglement between
Alice and Bob using Renyi entropies and also to {bound} the
underlying Hilbert space dimension. Second, when the {Hilbert space
dimension bound} is tight, the shared pure quantum state has to be
maximally entangled. Third, the second characterization gives a
sufficient condition that a Bell correlation cannot be generated by
particular quantum states. We also show that our results can be
generalized to the case of shared mixed states.

\end{abstract}
%\pacs{03.65.Aa, 03.65.Ud, 03.65.Wj}

\maketitle

{\em Introduction.---}In the study of quantum physics, frequently
the internal workings of a quantum device are not exactly known. For
example, it is often the case that we do not have sufficient
knowledge of the internal physical structure, or the precision of
the quantum controls is very limited, or even the devices we are
using cannot be trusted. In these cases, it could be that the only
reliable information available is the measurement statistics from
observing the quantum system. However, sometimes we still want to
draw nontrivial conclusions on the quantum properties of the
involved system. This sounds like a challenging, or even impossible
task, but it has been shown to be possible in many
cases~\cite{PR92,BMR92,MY98,MY04,BPA+08,GBHA10,CBB15,MBL+13,SVV15,SVV16,Ekert91,BHK05,AGM06,LBL+15}.
These kinds of tasks are called {\em device-independent} as their
application assumes only the correctness of quantum mechanics as a
valid description of nature, and is independent of the internal
workings of the devices used. Device-independence is a very valuable
property in physical implementations of various quantum schemes.
Typical examples of its usefulness include the transmission of
information safely using untrusted devices, and easy monitoring of
the overall performance of vulnerable quantum
devices~\cite{Ekert91,BHK05,AGM06,LBL+15}.

We consider in this {paper} the setting of a Bell experiment, i.e.,
two {spatially} separated parties sharing a quantum state and
performing local measurements on their subsystems. The corresponding
statistics of the measurement outcomes is called a Bell correlation.
It has been shown that the dimension and the entanglement of the
underlying quantum state can be quantified in a device-independent
way using only the Bell correlation
data~\cite{BPA+08,MBL+13,SVV15,SVV16}. In fact, some quantum states
can even be pinned down completely by their violations of particular
Bell inequalities, but this is only known to be possible for some
special cases~\cite{PR92,BMR92,MY98,MY04,YN13,Kaniewski16,WBMS16}.

In a Bell experiment, suppose a correlation is generated by
measuring {the} shared quantum state $\rho$. {We} often hope the
dimension of $\rho$ is as small as possible due to the fact that
quantum dimensionality is a precious resource. Interestingly, for an
arbitrary Bell correlation, it {is} known that this quantum state
with minimal dimension can always be pure~\cite{SVV15}.
Conveniently, a pure state can be described using its Schmidt
decomposition, where the Schmidt coefficients completely capture its
quantum properties.

In this {paper}, we give two device-independent characterizations of
the Schmidt coefficients of the state used in a general Bell
experiment. In particular, these characterizations are independent
of any Bell inequalities and are very easy to calculate using only
the correlation data. We show that these characterizations enjoy
various applications in many device-independent tasks. Concerning
the first characterization, we provide examples for which it is
tight and where the shared pure quantum states are actually pinned
down completely. Second, we show that it implies lower bounds on
both the dimension and the amount of entanglement of the underlying
quantum state, which are device-independent tasks that have drawn
much attention recently~\cite{BPA+08,SVV15,MBL+13}. We then show
that the second characterization allows us to exclude the pure state
that can produce a given Bell correlation from being particular
states. We also show that both of the characterizations can be
generalized to the case of shared mixed quantum states, where
Schmidt coefficients are replaced by eigenvalues of {the} reduced
density matrices of the two parties.

\begin{figure}[htbp]
   \centering
   \includegraphics[width=3in]{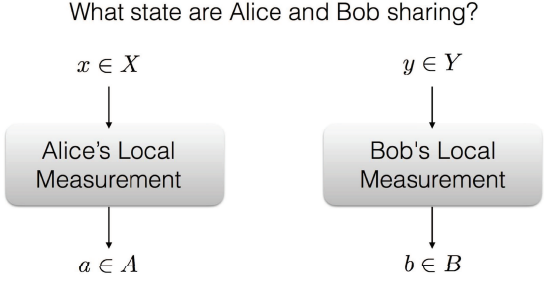}
   \caption{{Characterizing the quantum state in a Bell experiment with unknown internal workings.}
   }
   \label{fig:example}
\end{figure}

{\em Scenario.---}In a Bell scenario, the two separated parties,
Alice and Bob, share a pure quantum state $\ket{\psi}$ acting on
$\C^{d} \otimes \C^{d}$. They each have a local measurement
apparatus, and can choose different settings to measure their
respective subsystems. We denote the sets of the measurement
settings of Alice and Bob by $X$ and $Y$ respectively. For any $x\in
X$, the measurement $x$ is described as a local positive-operator
valued measure (POVM) ${\{ {M_{xa}} : a \in A \}}$, and similarly,
any measurement $y\in Y$ is described as {a POVM} $\{ {N_{yb}} : b
\in B \}$, where $A$ and $B$ are the sets of the measurement
outcomes of Alice and Bob respectively, {as illustrated in Fig.1.} A
Bell correlation $p$ is the collection of the joint conditional
probabilities $p(ab|xy)$ Alice and Bob observe, i.e.,
\begin{equation}\label{eq:schmidt}
%p(ab|xy) = \Tr[(M_{xa}\otimes N_{yb})\ket{\psi}\bra{\psi}].
{p(ab|xy) = \bra{\psi} M_{xa}\otimes N_{yb} \ket{\psi}.}
\end{equation}

Up to local change of bases, $\ket{\psi}$ can be Schmidt decomposed into the computational
basis as
\begin{equation} \label{purestate}
\ket{\psi}=\sum_{k=1}^d\sqrt{\lambda_k}\ket{k}\ket{k},
\end{equation}
where the Schmidt coefficients ${(\lambda_1, \ldots, \lambda_d)}$
are nonnegative. {Define} $D \equiv
\text{diag}\{\sqrt{\lambda_1},\sqrt{\lambda_2},...,\sqrt{\lambda_d}\}$.
It can be shown that
\begin{eqnarray}
\!\!\!\!\!\!\!\!\!\!\!\!\!\!\!\! p(ab|xy) & = & \Tr(M_{xa}\cdot DN_{yb}^*D)  = \Tr(D M_{xa} D \cdot N_{yb}^*) \label{eq:0}, \\
p(a|x) & = & \sum_b p(ab|xy) =\Tr(DM_{xa}D) \label{eq:11}, \\
p(b|y)& = & \sum_a p(ab|xy)=\Tr(DN^*_{yb}D) \label{eq:22},
\end{eqnarray}
where $S^*$ denotes the complex conjugate of the matrix~$S$.

\medskip
{\em A characterization of the Schmidt coefficients.---}For fixed
$y$ and $b$, if $p(b|y)\neq 0$, define the quantum state
\begin{equation} \label{defn}
\rho_{yb}\equiv \frac{1}{p(b|y)} DN^*_{yb}D,
\end{equation}
and notice that the probability that measurement $x$ outputs $a$,
when applied to $\rho_{yb}$, is given by $\frac{p(ab|xy)}{p(b|y)}$.

Now we want to estimate the distances between these quantum states.
For this, we utilize the fact that when two quantum states are
measured by the same measurement, the fidelity between two quantum
states $\rho$ and $\sigma$, defined {by} $\F(\rho, \sigma) := \|
\sqrt{\rho} \sqrt{\sigma} \|_1$, is upper bounded by that between
the probability distributions of measurement outcomes~\cite{NC00}.
Thus, for any $x\in X$, $y_1,y_2\in Y$, and $b_1,b_2\in B$, if
$p(b_1|y_1)>0$ and $p(b_2|y_2)>0$,
\begin{small}
\begin{equation*}
\F(\rho_{y_1b_1},\rho_{y_2b_2})\leq\sum_a
\sqrt{\frac{p(ab_1|xy_1)}{p(b_1|y_1)}\cdot\frac{p(ab_2|xy_2)}{p(b_2|y_2)}},
\end{equation*}
\end{small}
which means that
\begin{small}
\begin{equation}
\F(\rho_{y_1b_1},\rho_{y_2b_2})\leq\min_x\sum_a
\sqrt{\frac{p(ab_1|xy_1)}{p(b_1|y_1)}\cdot\frac{p(ab_2|xy_2)}{p(b_2|y_2)}}.
\end{equation}
\end{small} Combining this with the fact that $\Tr(\rho\sigma)\leq
\F(\rho,\sigma)^2$ for any quantum states $\rho$ and $\sigma$, we
obtain that
\begin{equation*}
\Tr(\rho_{y_1b_1}\rho_{y_2b_2})\leq \min_x \left(
\sum_a
\sqrt{\frac{p(ab_1|xy_1)}{p(b_1|y_1)}\cdot\frac{p(ab_2|xy_2)}{p(b_2|y_2)}}\right)^2.
\end{equation*}
Recalling the definition of $\rho_{yb}$, {we have}
\begin{small}
\begin{equation}\label{eq:1}
\Tr(N^*_{y_1b_1}D^2N^*_{y_2b_2}D^2)\leq \min_x \left(
\sum_a\sqrt{p(ab_1|xy_1)p(ab_2|xy_2)}\right)^2,
\end{equation}
\end{small}
%Note that \eqref{eq:1}
{which is also true when} $p(b_1|y_1)=0$ or $p(b_2|y_2)=0$.

On the other hand, for any $y \in Y$ it holds that $\sum_b D N^*_{yb} D =
D^2$ as $\{ N^*_{yb} : b \in B \}$ is a POVM. Thus, for any
$y_1,y_2 \in Y$, we have that
\begin{equation}\label{eq:2}
\sum_{i=1}^d \lambda_i^2 =
\Tr(D^4)=\sum_{b_1,b_2}\Tr(N^*_{y_1b_1}D^2N^*_{y_2b_2}D^2).
\end{equation}
Combining \eqref{eq:1} and \eqref{eq:2}, we obtain that
$\sum_i\lambda_i^2$ is upper bounded by
\begin{equation}\label{eq:onebound}
\min_{y_1,y_2} \sum_{b_1,b_2} \min_x \left(
\sum_a\sqrt{p(ab_1|xy_1)p(ab_2|xy_2)}\right)^2.
\end{equation}
Note that we {could} have regarded $\{N_{yb}\}$ as a measurement and
$DM^*_{xa}D/p(a|x)$ as a quantum state to {view} the correlation
data. {In this case,} by repeating the discussion above we conclude
that $\sum_i\lambda_i^2$ is also upper bounded by
\begin{equation}\label{eq:twobound}
\min_{x_1,x_2}\sum_{a_1,a_2}\min_y\left(\sum_b\sqrt{p(a_1b|x_1y)p(a_2b|x_2y)}\right)^2.
\end{equation}
Therefore, we have the following characterization for the Schmidt
coefficients.

\smallskip
\noindent \textbf{Theorem 1.} If a Bell correlation $p$ can be
generated by the state $\ket{\psi}$ with Schmidt coefficients
${(\lambda_1, \ldots, \lambda_d)}$, then
\begin{equation}\label{eq:main}
\sum_{i = 1}^d \lambda_i^2\leq\min\{ f_1(p),f_2(p)\},
\end{equation}
where $f_1(p)$ and $f_2(p)$ denote the values given in
\eqref{eq:onebound} and \eqref{eq:twobound}, respectively.
\smallskip

We now remark on Theorem 1. First, note that in the discussion
above, the dimension of the pure state can be arbitrary, thus
\eqref{eq:main} is valid for any pure state that generates $p$, not
just one of a particular dimension. For example, suppose
$\ket{\psi}$ is a quantum state generating some Bell correlation. We
can replace it with $\ket{\psi}\otimes\ket{\Phi}$ to produce the
same correlation, where $\ket{\Phi}$ is a redundant EPR pair shared
by Alice and Bob. It is easy to verify that for this new quantum
state, the sum of squares of the Schmidt coefficients has decreased,
which makes the bound \eqref{eq:main} looser. Therefore, Theorem 1
tends to provide a more meaningful result when the dimension of the
underlying system is close to minimal. We illustrate this in a later
example. {This also proves that one cannot hope to find a lower
bound on $\sum_i \lambda_i^2$ as a function of only the correlation
data.}

Second, we now consider the case when Alice and Bob share a mixed
state $\rho$. In this case, we can bound the \emph{purity} of
$\rho_{A}$ or $\rho_B$, where ${\rho_{A} \equiv \Tr_B(\rho)}$ and
${\rho_{B} \equiv \Tr_A(\rho)}$. The purity of a quantum state
$\rho$ is defined as $\Tr(\rho^2)$ (see~\cite{NC00}), and
$\Tr(\rho^2_{A})$ is precisely $\sum_{i=1}^d \lambda_i^2$ in the
case of the pure state \eqref{purestate}. To see how to bound the
purity of $\rho_A$, suppose Bob introduces a third subsystem $C$ on
his side to purify $\rho$ to be $\ket{\psi}_{ABC}$. Then by
performing an isometry, he maps his subsystem to a smaller one with
the same dimension as that of Alice {(seen to be possible by viewing
its Schmidt decomposition)}. Next, he adjusts the measurements he
uses by the same isometry. Then it can be verified that Alice and
Bob now have a {Bell experiment} that generates the same correlation
as before, where they share a pure quantum state on $\C^{d} \otimes
\C^{d}$ for some $d$. Note that Alice's reduced density matrix
remains unchanged in the whole process, and its eigenvalues are
exactly the Schmidt coefficients of the new pure state. Therefore
Theorem 1 gives an upper bound for the purity of $\rho_A$. Later we
discuss how this allows us to estimate the \emph{entanglement of
formation} of $\rho$.

{\em Several tight examples.---}To show that the bound
\eqref{eq:main} can be tight, we first consider an example in which
$A = B = X = Y = \{ 0, 1 \}$ and the correlation $p$ {is given} by
\begin{equation} \label{cor:chsh}
p(ab|xy) = \left\{
\begin{array}{rcl}
(2+\sqrt{2})/8, & \text{ if } & a \oplus b = {xy}, \\
(2-\sqrt{2})/8, & \text{ if } & a \oplus b \neq {xy},
\end{array}
\right.
\end{equation}
where $\oplus$ denotes the logical XOR of two bits. This correlation
corresponds to the optimal strategy for the CHSH game~\cite{CHSH69}
and can be generated by the maximally entangled state in $\C^2
\otimes \C^2$.
From \eqref{eq:main} we can see that $\sum_{i = 1}^d \lambda_i^2
\leq 1/2$, which is tight.

For a second example, we now apply our bound to an extreme point of
the no-signaling polytope in the setting $|X| = |Y| = |A| = |B| = 3$
(see Table III of \cite{JM05}). We find that $f_1(p) = 0$ (seen by
choosing $y_1 = 0, y_2 = 2$). Thus, $\lambda_i = 0$ for every $i$,
implying no finite-dimensional quantum state exists which generates
this correlation. Thus, we can certify the non-quantumness of
particular correlations.

As the last example, we set $X = Y = \{ 1, 2, 3 \}$ and ${A = B = \{
0, 1 \}^3}$ and consider the correlation
\begin{eqnarray} \label{eq:magic}
& & \!\!\!\!\!\!\!\!\! p(ab|xy) \! = \! \left\{
\begin{array}{cl}
\! {1/8}, & \text{if } a_y = b_x, \; a \text{ has even parity},  \\
\! \phantom{1/8} & \phantom{\text{if }} \text{\ \ \ and } b \text{ has odd parity}, \\
\! 0, & \text{otherwise}. \\
\end{array}
\right.
\end{eqnarray}
This correlation is optimal for the Magic Square
Game~\cite{Mermin90,Peres90,Aravind02}, and can be generated if
Alice and Bob share the maximally entangled state in $\C^4 \otimes
\C^4$.

We now ask the question whether it is possible to generate this
correlation with any other pure state of the same dimension. The
answer is no, and we can prove this using Theorem 1. By
straightforward calculation, it can be shown that the right side of
\eqref{eq:main} for this case is $1/4$, which again is tight.
Moreover, for a pure state on $\C^{4} \otimes \C^{4}$, the minimum
value of $\sum_i\lambda_i^2$ is $1/4$, and it can be achieved only
by a maximally entangled state. Therefore, in this case, Theorem 1
certifies that the pure quantum state on $\C^{4} \otimes \C^{4}$
that can generate \eqref{eq:magic} is unique up to local unitary
transformations. Actually, even if we allow the shared state to be
mixed, it has been shown in a recent work~\cite{WBMS16} that the
state must still be maximally entangled on $\C^4 \otimes \C^4$.
These results are useful in the line of research known as
\emph{self-testing}~\cite{PR92,BMR92,MY98,MY04}. Note that a similar
analysis can be applied to the correlation \eqref{cor:chsh}.

\medskip
{\em Relation to device-independent dimension
test.---}Device-independent lower bounds on the dimension of a
quantum state used in a Bell setting is a very interesting problem
that has attracted much attention recently~\cite{BPA+08,SVV15}.
Recall that for any Bell correlation, a quantum state with the
minimal size that produces this correlation can always be
pure~\cite{SVV15}. In our notation, if $\ket{\psi} \in \C^d \otimes
\C^d$ generates the correlation $p$, we would like to lower bound
$d$ using only the {correlation} data. Noting that $d \geq
1/(\sum_{i =1}^d \lambda_i^2)$ is valid for any pure state,
\eqref{eq:main} {immediately implies the two \emph{lower bounds} for
the underlying} Hilbert space dimension
\begin{small}
\begin{eqnarray}
\!\!\!\!\!\!\!\!\!\! \bigg( \min_{y_1,y_2} \bigg( \sum_{b_1,b_2} \min_x \sum_a \sqrt{p(ab_1|xy_1) p(ab_2|xy_2)} \bigg)^2 \bigg)^{-1} \label{eq:old1}, \\
\!\!\!\!\!\!\!\!\!\! \bigg( \min_{x_1,x_2} \bigg( \sum_{a_1,a_2} \min_y \sum_b \sqrt{p(a_1b|x_1y) p(a_2b|x_2y)} \bigg)^2 \bigg)^{-1}
\label{eq:old2},
\end{eqnarray}
\end{small}
recovering the main result in Ref.~\cite{SVV15}. However, these
lower bounds on the dimension do not imply our result Theorem 1.

\medskip
{\em Quantification of entanglement.---}Since quantum properties of
a bipartite pure quantum state are captured completely by the
Schmidt coefficients, our bound \eqref{eq:main} can be used to
characterize other properties as well in a device-independent
manner. As a natural application, we now consider quantifying the
amount of entanglement shared by Alice and Bob.

For this, we first recall that the generalized Renyi entanglement
entropies of a mixed state $\rho$ are defined as
\begin{equation}
S_n(\rho) \equiv \frac{1}{1-n}\log \big( \Tr(\rho^n) \big),
\end{equation}
where $n>0$ is a real number. It can be shown that $S_n$ is a
non-increasing function in $n$ which, as $n$ approaches $1$,
converges onto the well-known von Neumann entropy
\begin{equation}
S(\rho) \equiv -\Tr(\rho\log(\rho)).
\end{equation}
As a result, $S_2(\rho)$ is a natural lower bound for the von
Neumann entropy $S(\rho)$. Revisiting Theorem 1, if a pure state
generates a correlation $p$, it is clear that we can bound
$S_2(\rho_A)$, where $\rho_A$ is Alice's reduced density matrix, as
\begin{equation} \label{entropy}
S(\rho_A)\geq S_2(\rho_A)\geq-\log \big(\min\{ f_1(p),f_2(p)\}
\big).
\end{equation}
Note again that a lower bound on the dimension does not directly
imply any lower bounds on the entropy. On the other hand, for a
fixed Bell correlation $p$, there does not exist a general upper
bound on $S(\rho_A)$ since Alice and Bob can always carry redundant
{EPR pairs} and still generate the same correlation.

As an example, we now use \eqref{entropy} to consider the $I3322$
Bell inequality~\cite{Froissart81}, which is quite interesting as
numerical {evidence} suggests that to violate this Bell inequality
maximally, infinite-dimensional Hilbert spaces are
required~\cite{PV10}. By applying \eqref{entropy} to a Bell
correlation produced by a quantum state in $\C^{49} \otimes \C^{49}$
that approximates the maximal violation given in Ref.\cite{PV10}, we
obtain that the von Neumann entanglement entropy needed to produce
this correlation from a shared pure state is at least $0.67$.

Since in practical experiments quantum states are {often} mixed, we
next {briefly discuss the case when}
%we suppose that in a Bell experiment,
the shared state $\rho$ is unknown {but assumed to be close to pure,
i.e.,   that $\Tr(\rho^2)>1-\eta$, where $\eta$ is a small positive
number.
%, {meaning} that $\rho$ is close to pure.
Note that with this assumption, it is not  completely
device-independent any longer. However, this is still a realistic
setting due to the remarkable improvements in quantum
experimentation in recent years}. We now show that our results allow
us to estimate the \emph{entanglement of formation} of $\rho$,
{denoted by $E_f(\rho)$ and defined to be
\begin{small}
\begin{equation}
E_f(\rho)\equiv \min{\sum_ip_iS(\rho_{i})},
\end{equation}
\end{small}
where the minimum is taken over all ensembles
$\{p_i,\ket{\alpha_i}\}$ generating $\rho$, and
$\rho_{i}=\Tr_B(\ket{\alpha_i}\bra{\alpha_i})$}. Suppose an
orthogonal decomposition of $\rho$ is
$\rho=\sum_{i=1}^ka_k\ket{\psi_i}\bra{\psi_i}$, where $a_i\geq a_j$
for $i<j$. Then it can be shown that
\begin{equation}
a_1\geq\frac{1}{2}+\sqrt{\frac{1}{2}\left(\frac{1}{2}-\eta\right)}\approx1-\frac{1}{2}\eta.
\end{equation}
Thus the distance between $\rho$ and $\ket{\psi_1}\bra{\psi_1}$ is
small. {Also, we have}
$\Tr(\rho^2_{A1})\leq\frac{1}{a_1^2}\Tr(\rho_A^2)$, where
$\rho_{A}=\Tr_B(\rho)$ and
$\rho_{A1}=\Tr_B(\ket{\psi_1}\bra{\psi_1})$. Combining this fact
with the upper bound for $\Tr(\rho_A^2)$ mentioned above, one can
lower bound the entanglement entropy of $\ket{\psi_1}\bra{\psi_1}$,
which is also its entanglement of formation
$E_f(\ket{\psi_1}\bra{\psi_1})$. {Lastly, according to the
continuous property of the {entanglement of
formation}~\cite{Nielsen00}, it holds that
\begin{equation}
|E_f(\rho)-E_f(\ket{\psi_1}\bra{\psi_1})|\leq
\sqrt{2\eta}(9\log(d)-\log(2\eta)).
\end{equation}
This way one can obtain a lower bound for $E_f(\rho)$}.

\medskip
{\em The smallest Schmidt coefficient.---}In this section, we give
another necessary condition that the set of Schmidt coefficients
must satisfy. Suppose we define $\lambda_{\min}$ as the least
nonzero Schmidt coefficient of the pure state that generates a
correlation $p$. We now show that it can be upper bounded in a
device-independent manner by a function of the correlation data.

Using the isometry argument mentioned before, we can assume without
loss of generality that the number of nonzero Schmidt coefficients
is $d$, i.e., the shared pure state {$\ket{\psi} \in \C^d \otimes
\C^d$} has full Schmidt rank. Note that for any positive
semidefinite matrices $A$ and $B$, we have that $\Tr(AB) \leq
\Tr(A)\Tr(B)$. Then using \eqref{eq:0}, we have
\begin{eqnarray}
p(ab|xy) & \leq & \Tr(M_{xa})\cdot p(b|y), \\
p(ab|xy) & \leq & \Tr(N_{yb})\cdot p(a|x).
\end{eqnarray}
By \eqref{eq:11} and \eqref{eq:22} we have
\begin{equation}
\Tr(M_{xa}) \leq \frac{p(a|x)}{\lambda_{\min}} \quad  \text{ and } \quad
\Tr(N_{yb}) \leq \frac{p(b|y)}{\lambda_{\min}}.
\end{equation}

Considering that these {inequalities} are valid for any choice of
parameters, we obtain the following theorem.

\smallskip
\noindent \textbf{Theorem 2.} If a Bell correlation $p$ can
be generated by the state $\ket{\psi}$ with least nonzero Schmidt coefficient $\lambda_{\min}$, it
holds that
\begin{equation}\label{eq:min}
\lambda_{\min} \leq \min_{x,y,a,b} \frac{p(a|x)p(b|y)}{p(ab|xy)}. %{\;\text{ when } \; p(ab|xy) \neq 0}.
\end{equation}

We now comment on how Theorem 2 can be tight. As an example,
consider the BB84 correlation defined as
$p(ab|xy)=\frac{1+ab\delta_{xy}}{4}$~\cite{GBS16}, %\snote{I doubt this
%is where it first appeared. We should have a look. Also, is there a
%reason it is not in a display environment?}
where $a,b,x,y\in\{-1,1\}$. This Bell correlation can be generated
by the maximally entangled state in $\C^2 \otimes \C^2$. A quick
calculation of \eqref{eq:min} shows that $\lambda_{\min} \leq 1/2$,
which is tight.
%$\ket{\Phi^+}=\frac{1}{\sqrt{2}}(\ket{00}+\ket{11})$, and the bound
%given by \eqref{eq:min} is $1/2$.

Again, if Alice and Bob share a mixed state $\rho$, Theorem 2 can be
used to upper bound the minimum nonzero eigenvalues of $\rho_A$ and
$\rho_B$.

One may ask whether we can lower bound the greatest Schmidt
coefficient based only on the correlation data. It turns out that it
is not possible. Again, for any Bell experiment, if Alice and Bob
introduce a redundant pure state, the greatest Schmidt coefficient
can become arbitrarily small while still generating the same
correlation.

Above we have seen examples where pure states {of certain}
dimensions {which generate particular} Bell correlations have to be
{maximal entangled}. We now use Theorem 2, in the opposite manner,
to show that a correlation \emph{cannot} be generated using a
particular state, again under dimension assumptions. For this,
suppose that $p$ is generated {by $\ket{\psi} \in \C^d \otimes
\C^d$}. Then if \eqref{eq:min} certifies that ${\lambda_{\min} <
1/d}$, we can conclude that, independent of the local measurements
Alice and Bob may apply, $\ket{\psi}$ cannot be maximally entangled.
In other words, $p$ cannot be {reproduced} by any maximally
entangled state of local dimension up to $d$. Of course, this can be
used to rule out other states as well, depending on the dimension
and bound on $\lambda_{\min}$.

We now illustrate this with a concrete example. Suppose Alice and
Bob fix {some choice} of measurements $x$ and $y$, and each
measurement has three outcomes $\{ 1,2,3 \}$. We specify some of the
probabilities in a possible correlation $p$ below:
\begin{equation*}
\begin{bmatrix}
1/10 & 1/100 & 1/100\\
1/100 & * & *\\
1/100 & * & *
\end{bmatrix},
\end{equation*}
where the $(a,b)$-entry is $p(ab|xy)$, and the {asterisks represent
unspecified probabilities}. According to Ref.~\cite{JSWZ13}, the
minimum size of quantum state that can generate {such a {partial}
correlation} $p$ {has local dimension} at most $3$. Meanwhile, it
can be verified using \eqref{eq:min} that $\lambda_{\min} \leq
18/125$, which is {strictly} less than $1/3$. Therefore, it is clear
that any pure state {in $\C^3 \otimes \C^3$} which generates $p$
cannot be maximally entangled. In fact, we would require a maximally
entangled state to have local dimension of at least $7$ to generate
$p$. Furthermore, if we restrict to a state {in $\C^2 \otimes
\C^2$}, then {such a correlation} cannot be generated by any state
of the form $\sqrt{a} \ket{00} + \sqrt{1-a} \ket{11}$ where $a \in
(18/125, 107/125)$.

\medskip
{\em Conclusions.---}For an arbitrary Bell correlation produced by
locally measuring a bipartite pure quantum state, we have given two
characterizations for its Schmidt coefficients, which can be
generalized to the case of shared mixed states. Also, we showed that
they have various applications in many device-independent quantum
processing tasks. Since our bounds only involve simple functions of
the Bell correlation data, they are quite robust against errors in
statistical data, making them usable in practical quantum tasks. We
hope these results will lead to more nontrivial applications in
quantum physics and quantum information theory, and particularly, we
hope the entanglement quantification application can be helpful in
future quantum experiments.

\medskip
\begin{acknowledgments}
We thank K\'aroly F. P\'al and Tam\'as V\'ertesi for sending us
numerical data {on the $I3322$ inequality} and Koon Tong Goh for
helpful discussions. Z.W. is supported by the Singapore National
Research Foundation under NRF RF Award No.~NRF-NRFF2013-13. Research
at the Centre for Quantum Technologies is partially funded through
the Tier 3 Grant ``Random numbers from quantum processes,''
(MOE2012-T3-1-009).
\end{acknowledgments}

\end{document}